\documentclass[reprint,
superscriptaddress,
 twocolumn,
 nofootinbib,
 nobibnotes,
 longbibliography,
 amsmath,amssymb,
 aps,
prb,
]{revtex4-2}
\pdfoutput=1
\usepackage{physics}
\usepackage{amsmath}
\usepackage{xcolor}
\usepackage{float}
\usepackage{graphicx}
\usepackage{dsfont}
\usepackage{comment}
\usepackage{ulem}
\usepackage{soul}
\usepackage[colorlinks=true,citecolor=blue,urlcolor=blue,linkcolor=blue]{hyperref}

\RequirePackage{stix}

\newcommand{\figref}{Fig. \ref}
\newcommand{\secref}{Sec. \ref}
\newcommand{\appendixref}{Appendix \ref}
\renewcommand{\eqref}[1]{{Eq.~(\ref{#1})}}

\usepackage{bm}

\renewcommand{\Vec}[1]{\bm{#1}}
\newcommand{\inter}{{\rm{int}}}

\newcommand{\BZ}{{\rm{BZ}}}
\newcommand{\MF}{{\rm{MF}}}
\newcommand{\FD}{{\rm{FD}}}
\newcommand{\ii}{{\rm{i}}}
\newcommand{\ee}{{\rm{e}}}
\newcommand{\singtrip}{{{\rm{s}} / {\rm{t}}}}
\newcommand{\unitary}{{U}}

\begin{document}
\title{Superconductivity from altermagnetic fluctuations: the crucial role of sublattice polarization}

\author{Mats Barkman}
\affiliation{Niels Bohr Institute, University of Copenhagen, DK-2200 Copenhagen, Denmark}

\author{Andreas Kreisel}
\affiliation{Niels Bohr Institute, University of Copenhagen, DK-2200 Copenhagen, Denmark}
\affiliation{Department of Physics and Astronomy, Uppsala University, Box 524, 751 20 Uppsala, Sweden}

\author{Brian M. Andersen}
\affiliation{Niels Bohr Institute, University of Copenhagen, DK-2200 Copenhagen, Denmark}

\begin{abstract}
Unconventional superconductivity in correlated systems often appears in the vicinity of a magnetically ordered state. 
This situation is also relevant for altermagnets, characterized by a nonrelativistic spin-split electronic structure in the presence of N\'eel order. 
Many proposed altermagnetic candidates, however, never cross the magnetic instability. 
We study superconductivity arising from altermagnetic fluctuations in such systems and show that the sublattice polarization (SLP) of the normal-state bands plays a pivotal role in determining the superconducting state. 
SLP, an inherent property of altermagnetic band structures that persists even without magnetic order, prevents Cooper pairs from forming between different sublattices. 
This suppresses pairing mediated by the nearest-neighbor antiferromagnetic exchange, which favors spin-singlet superconductivity, and leaves longer-range same-sublattice bonds dominant.
When the next-nearest-neighbor exchange is ferromagnetic, as expected for frustration-free N\'eel order, this results in spin-triplet superconductivity. 
This mechanism is fundamentally different from that in altermagnets with long-range N\'eel order, where the spin-split Fermi surfaces favor equal-spin pairing. 
Finally, when longer-range interactions in the disordered phase instead favor singlet pairing, SLP leads to singlet states with characteristic properties distinct from nearest-neighbor pairing states.

\end{abstract}

\maketitle

\section{Introduction}

The unique properties of altermagnets are currently being investigated both for technological purposes and for new phenomena within basic research. As opposed to antiferromagnetism, altermagnetism does not imply period doubling in real space. Rather it consists of intra-unit-cell Néel ordering of collinear moments on sublattice sites that are not related by inversion or sub-unit-cell translations, but proper or improper rotations~\cite{Hayami2019,Ahn2019,Smejkal2020Jun,Yuan2020Jul,Mazin2021,Smejkal_Emerging_2022,Smejkal_Conventional_2022}. This distinct symmetry classification of altermagnets has important implications for their electronic band structure, most importantly an exchange-generated nonrelativistic momentum-dependent spin splitting. Microscopic theories able to capture the origin of the altermagnetic spin-splitting directly from the underlying lattice symmetries and the Néel ordering must include both intra-unit-cell sublattice and spin degrees of freedom. Such minimal models necessarily feature nontrivial quantum geometric properties with associated important sublattice polarized bands~\cite{Roig_Minimal_2024,Antonenko,Roig_Landau,Jannik2025,Heinsdorf}. We stress that sublattice polarization (SLP) is an inherent fundamental property of crystallographic altermagnets. In this study we show that SLP has crucial consequences also for the ordered states that emerge from altermagnetic fluctuations. 

Altermagnetic metals may become superconducting at low temperatures. The nature of the preferred superconducting ground state induced by the altermagnetic spin splitting has been extensively discussed theoretically~\cite{Bose2024,Hong2025,Brekke_Twodimensional_2023,Carvalho2024,Chakraborty2024,Leraand2025,Parthenios,Zhu2023,Li2024,Ghorashi2024,Khodas2025,Parshukov2025m,Monkman}. The resulting system is a new platform for studying the interplay between unconventional magnetism and superconductivity. For example, several works have pointed out equal-spin-paired (ESP) $p$-wave triplet states as the preferred superconducting phase in the limit of large altermagnetic spin splitting~\cite{Zhu2023,Hong2025,Monkman,Hestbjerg,Zou2026}. Such ESP condensates feature spin-polarized supercurrents and topological surface states~\cite{Zhu2023,Monkman}. 
Other works have suggested the possibility of finite-momentum pairing ~\cite{Chakraborty_Zerofield_2023, Zhang2024, Hong2025}. 

Another pathway to interesting superconducting ground states emerging from the interplay with altermagnetism refers to superconducting states arising from a failed altermagnet, i.e. a material that could become altermagnetic, yet superconductivity condenses and the magnetic order never sets in~\cite{Wu2025,Lu_fluc_2025,Ma_fluc_2025}. Indeed, many known unconventional superconductors are proximate to ordered magnetic phases, and comply with the general notion of being generated by  magnetic fluctuations~\cite{Scalapino2012,RomerRMP}.  In this respect, a recent theoretical study proposed that the superconductivity observed in MnB$_4$ under high pressure is driven by altermagnetic fluctuations~\cite{Radevych}. Solving the pairing problem, the leading instability is found to be an extended $s$-wave state, with the triplet channel non-competitive. This is in contrast to Ref.~\cite{Wu2025}, where an altermagnet on the Lieb lattice was studied in the disordered phase close to a putative altermagnetic quantum critical point. There the range of the fluctuations decides the outcome: long-wavelength fluctuations act much like ferromagnetic ones and stabilize a conventional spin-triplet $p$-wave state, whereas shorter-range fluctuations favor intra-unit-cell pairing, of $s$-, $p$- or $d$-wave symmetry depending on the shape of the Fermi surface. Triplet pairing from altermagnetic fluctuations has also been reported for a Hubbard model on a $\sqrt{5}\times\sqrt{5}$ vacancy-ordered lattice, where singlet pairing instead prevails whenever conventional antiferromagnetic fluctuations dominate~\cite{Ma_fluc_2025}, and for a two-orbital model near van Hove filling, where the broken inversion symmetry of the
fluctuations favors the triplet channel~\cite{Lu_fluc_2025}. Which channel prevails thus depends on the range of the fluctuations, the orbital content and the Fermi surface. What has not been addressed, however, is the role played by the SLP of the normal-state bands, a property inherent to altermagnets that survives even when the magnetic order never develops. The distinction matters since the intra-unit-cell states of Ref.~\cite{Wu2025}, for example, are built from electrons residing on different
sublattices, whereas strong enough SLP acts to block such pairs. As we show below, it is this suppression of the nearest-neighbor (NN) channel, rather than the form of the fluctuation spectrum, that can leave longer-range same-sublattice bonds, and thereby spin-triplet and intra-unit-cell sign-modulated order, dominant.

In this work we study the preferred superconducting states arising out of a failed altermagnet starting from a generic minimal model for altermagnetism. 
We address and draw attention to features unique to altermagnetic fluctuations, i.e. features that are typically absent in materials with either ferromagnetic or antiferromagnetic fluctuations.
Antiferromagnetic fluctuations result in spin-singlet superconductivity dominated by NN bonds~\cite{Scalapino2012,RomerRMP,RomerAF}.
In contrast, NN pairing is suppressed by altermagnetic fluctuations as a result of the SLP of the normal state. Indeed it is well known, particularly from studies of correlated kagome systems, that SLP can severely affect the properties of the emergent phases at low temperatures~\cite{KieselEA12,Kiesel2013Unconventional,Yu2012,Wang2013,Romer2022,Holbaek_Unconventional_2023,YiDai2024,Sofie2026}. In the present case, the resulting leading superconducting order, now supported by longer-range bonds, may significantly differ from the NN dominated pairing.
For example, the longer-range pairing can lead to spin-triplet superconductivity and intra-unit-cell sign-modulated order.
We demonstrate this effect using two approaches, first using a Heisenberg exchange interaction and second using an RPA approach where a dressed onsite Hubbard repulsion results in effective pairing for superconductivity.

\section{Methodology}

We model altermagnetism using a tight-binding formalism, where we use fermionic creation and annihilation operators with both spin and sublattice degrees of freedom. 
Let $c_{\sigma a}$ denote the annihilation operator of an electron with spin $\sigma \in \lbrace \uparrow, \downarrow\rbrace$ and sublattice degree of freedom $a \in \lbrace {\rm{A}}, {\rm{B}} \rbrace$ and introduce $\Vec{c}^\dagger = 
\begin{pmatrix}
    c_{\uparrow {\rm{A}}}^\dagger & c_{\uparrow {\rm{B}}}^\dagger
    & c_{\downarrow {\rm{A}}}^\dagger & c_{\downarrow {\rm{B}}}^\dagger
\end{pmatrix}
$.
In reciprocal space, the normal state minimal model Hamiltonian  can be written
\begin{equation}
    H_0 =  \int_{\textbf{k} \in \BZ} \frac{\dd ^d \textbf{k}}{|\Omega_\BZ|}
    \Vec{c}^\dagger (\textbf{k}) \mathcal{H}_0 (\textbf{k}) \Vec{c}(\textbf{k}), \label{eq_normal_state}
\end{equation}
where integration is carried out over the first Brillouin zone (BZ) ($|\Omega_\BZ|$ is the volume of the first BZ) and
\begin{equation} \label{eq: minimal model Hamiltonian}
    \mathcal{H}_0 (\textbf{k}) =
    \sigma_0 \otimes [\epsilon_0 (\textbf{k}) \tau_0 + (\Vec{\epsilon} (\textbf{k}) \cdot \Vec{\tau})]
    +
    (\Vec{N} \cdot \Vec{\sigma}) \otimes \tau_z,
\end{equation}
where $\sigma_\mu$ ($\tau_\mu$) are the Pauli matrices acting on the spin (sublattice) degree of freedom. 
The first term, defined by $\epsilon_\mu (\textbf{k})$, describes the normal state in the absence of magnetic ordering.
$\epsilon_x$ and $\epsilon_y$ correspond to inter-sublattice terms, while $\epsilon_0$ and $\epsilon_z$ correspond to intra-sublattice terms, where $\epsilon_z$ sets the difference between the two sublattice degrees of freedom.
The second term contains the magnetic N\'eel order parameter $\Vec{N}$.
Together with $\epsilon_z (\textbf{k})$ (i.e. the term in the normal state Hamiltonian that differs between the two sublattice degrees of freedom), the presence of $\Vec{N}$ results in the spin-split dispersions characteristic of altermagnets~\cite{Roig_Minimal_2024,Roig_Landau,Jannik2025}. 

\begin{figure}
    \centering
    \includegraphics[width=1.05\linewidth]{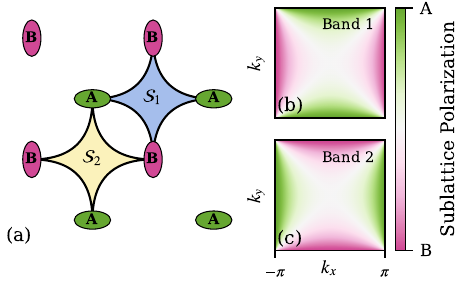}
    \caption{
    Illustration of the Lieb lattice (a) and the sublattice polarization of the two bands (b-c).
    The unit cell contains two sublattices (A and B) and two symmetry-distinct NNN hoppings $t_\pm =t_2 \pm t_z $ (across the plaquettes $\mathcal{S}_1$ and $\mathcal{S}_2$ respectively) in addition to the NN hopping $t_1$.
    The sublattice polarization increases with $t_z$.
    }
    \label{fig: illustration sublattice}
\end{figure}

Here, we are interested in the case of altermagnetic fluctuations, where no magnetic order has developed (that is $\Vec{N} = \Vec{0}$).
Importantly, the normal state still contains the sublattice degree of freedom and has two distinct spin-degenerate bands. 
A characteristic feature of these two bands is that they are sublattice polarized. 
This means that the Bloch wavefunctions have unequal weights on the two sublattice degrees of freedom depending on the reciprocal space coordinate $\textbf{k}$.
For example, the SLP of the two bands in the two-dimensional (2D) Lieb lattice (sometimes also referred to as the \textit{inverse} Lieb lattice) is shown in \figref{fig: illustration sublattice}.
In real space, the origin of the SLP is an intra-sublattice hopping term that differs between the $x$- and $y$-direction. Four-fold rotational symmetry enforces that the horizontal A-hopping is the same as the vertical B-hopping and vice versa.
This results in $\epsilon_z( \textbf{k}) = -2t_z (\cos k_x - \cos k_y)$ with nodes at $k_x = \pm k_y$, which are inherited by the SLP.
Importantly, SLP is present even in the absence of magnetic order. For other classes of 2D minimal models relevant for altermagnets, we refer to Ref.~\cite{Jannik2025}.

In this study, we investigate the consequences of SLP on the superconducting instabilities in the presence of altermagnetic fluctuations.
Superconductivity is driven by an interaction Hamiltonian $H_\inter$, which would also be the origin of the magnetic N\'eel order parameter in the normal state Hamiltonian. However, for the case of failed altermagnets, we carry out a mean-field decoupling only in the Cooper channel, considering the corresponding superconducting instabilities.
In the following sections we elaborate on the specific interaction Hamiltonian used, and the results demonstrating the unique features of the superconducting state arising in the presence of strong SLP.

\section{Results}

\subsection{Heisenberg interaction}

We consider an interaction Hamiltonian with both onsite Hubbard repulsion and extended Heisenberg exchange couplings given by,
\begin{equation} \label{eq: Heisenberg interaction}
    H_\inter = U \sum_{\textbf{r}} n_\uparrow (\textbf{r}) n_\downarrow (\textbf{r})  
    +
    \sum_{\textbf{r} \textbf{r}'} J_{\textbf{r} \textbf{r}'} \Vec{S}(\textbf{r}) \cdot \Vec{S}(\textbf{r}'),
\end{equation}
where $n_\sigma (\textbf{r})= c^\dagger_\sigma (\textbf{r}) c_\sigma (\textbf{r})$ and $U>0$ is the onsite Hubbard repulsion, and the sum in the Heisenberg term runs over bonds $\textbf{r} \textbf{r}'$ and $S_i(\textbf{r}) = \Vec{c}^\dagger (\textbf{r}) \sigma_i \Vec{c} (\textbf{r})$. 
We are interested in the case in which the exchange interactions $J_{\textbf{r} \textbf{r}'}$ favor a staggered N\'eel order, but the system instead develops superconducting order.
For a given bond, recall that
\begin{align}
    \Vec{S}(\textbf{r}) \cdot \Vec{S}(\textbf{r}') \ket{\text{singlet}} & = -3 \ket{\text{singlet}}, \\
    \Vec{S}(\textbf{r}) \cdot \Vec{S}(\textbf{r}') \ket{\text{triplet}} & = +1\ket{\text{triplet}},
\end{align}
which is key to understanding which superconducting order is favored. Interaction terms which favor anti-parallel spins ($J_{\textbf{r} \textbf{r}'} > 0$) promote singlet superconductivity, while terms that favor parallel spins ($J_{\textbf{r} \textbf{r}'} < 0$) promote triplet superconductivity. A recent {\it ab initio} study of Lieb lattice materials computed the exchange couplings for several different material candidates~\cite{Chang2026}. Interestingly, while the NN exchange was always found to be antiferromagnetic, the NNN exchange coupling can either be ferromagnetic or antiferromagnetic~\cite{Chang2026}. Exchange couplings have also been computed for RuO$_2$ in relation to the discussion of chiral altermagnetic magnon modes~\cite{Smejkalmagnons} giving similar dominant antiferromagnetic couplings. Below, we explore parameter regimes of exchange couplings qualitatively compatible with these first-principles studies.
We apply a mean-field approximation to the interaction in \eqref{eq: Heisenberg interaction} in the Cooper channel. Assuming translation invariance, we obtain a self-consistency equation for the superconducting gap function which we linearize to obtain the linearized gap equation (LGE) given by
\begin{equation} \label{eq: LGE}
    \mathcal{J} \mathcal{M} \Vec{\Delta}= \lambda \Vec{\Delta},
\end{equation}
where the components of the vector $\Vec{\Delta}$ correspond to the gap function at different bonds, $\mathcal{J}$ is a diagonal matrix containing the Heisenberg exchange couplings and Hubbard repulsion $U$, and $\mathcal{M}$ is the matrix obtained from linearizing the self-consistency equation. 
The details of deriving the mean-field equations and \eqref{eq: LGE} are shown in \appendixref{appendix: LGE derivation}. 
We determine the leading superconducting instability by analyzing the eigenvalues and eigenvectors of $\mathcal{JM}$, where the largest eigenvalue $\lambda_{\max}$ determines the leading superconducting order. 
We restrict ourselves, without loss of generality, to opposite-spin pairing, since in the absence of magnetic order and spin-orbit coupling, all triplet states are degenerate.

Importantly, $\mathcal{M}$ depends on the electronic band structure and the unitary transformation between sublattice and band degrees of freedom. 
Consequently, the SLP plays an important role when determining the preferred superconducting order. 
Let us briefly explain how we vary the SLP. 
Consider the Lieb lattice (which we will focus on in this study).
As previously mentioned, the SLP is determined by the term $\epsilon_z(\textbf{k}) ~= ~-2t_z (\cos k_x - \cos k_y)$ in the normal state Hamiltonian.
We can therefore increase the SLP by increasing $t_z$.
Changing $t_z$ of course also changes the dispersion relation and the Fermi surface shape, which in turn can affect the favored superconducting order.
We want to isolate the effect of SLP on the preferred superconducting order. The strategy is to keep the Fermi surface fixed (i.e. the Fermi surface for $t_z = 0$) upon calculating the LGE matrix $\mathcal{M}$, while varying the SLP on that fixed Fermi surface.
By doing so we eliminate any effects Fermi surface reshaping has on the preferred superconducting order, focusing only on the effects from changes in SLP.

For simplicity we restrict ourselves to NN exchange $J_1$ and next-nearest-neighbor (NNN) exchange $J_2$, in addition to an onsite Hubbard repulsion $U$. We assume that the Hubbard repulsion $U$ is much stronger than the Heisenberg exchange coupling. Note that one could allow for two symmetry-distinct NNN exchange couplings $J_\pm = J_2 \pm J_z$ (similar to the corresponding hoppings). In order to focus on the effects from the SLP, we here set $J_z = 0$.

Before going into concrete examples, let us outline the main mechanism behind the preferred superconducting order induced by altermagnetic fluctuations and SLP.
Let us start with an antiferromagnet, that is $t_z=0$ and we consider some exchange interactions $J_1$ and $J_2$ such that staggered magnetic order is preferred.
The NN coupling $J_1 > 0$ is dominant, which favors singlet superconductivity. Now consider instead an altermagnet: even in the absence of magnetic order, there is a nontrivial band structure characterized by SLP in the BZ.
That is, in certain regions of momentum space, the electrons in each band are predominantly located at one or the other sublattice.
If the Fermi surface is located where the SLP is strong, this severely prevents the formation of Cooper pairs consisting of electrons of different sublattices, effectively suppressing the NN coupling $J_1$. 
This makes the NNN coupling dominant, which can support fundamentally different superconducting orders compared to the NN coupling. For example, if $J_2 < 0$ this can result in a leading spin-triplet state. 

\subsubsection{Case 1: Leading spin-triplet superconductivity}

\begin{figure}
    \centering
    \includegraphics[width=1.0\linewidth]{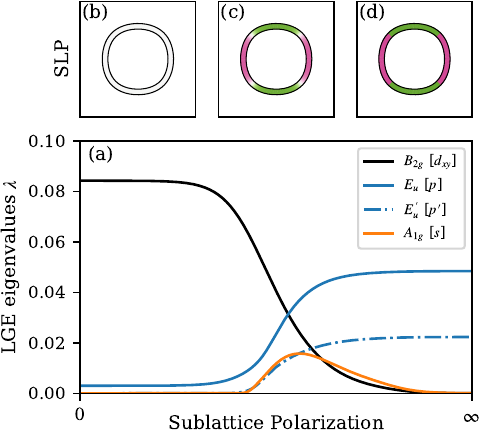}
    \caption{ (a) Eigenvalues of the linearized gap equation as SLP increases. The upper panels (b-d) show the SLP on the fixed Fermi surface. The NN coupling is $J_1 = 1$, which favors singlet superconducting order, while NNN coupling is $J_2 = -0.25$, favoring triplet superconductivity. For weak SLP, the NN pairing dominates, favoring singlet superconductivity. When the SLP is strong, the dominant NN pairing is suppressed, resulting in a decrease in the LGE eigenvalues and the emergence of a leading spin-triplet superconducting order.
    } 
    \label{fig: lieb main example}
\end{figure}

Let us illustrate with a concrete example how the suppression of the NN singlet channels leads to an instability toward longer-range triplet superconductivity.
In order to favor staggered magnetism without competing NN and NNN couplings, we should choose $J_1 > 0$ and $J_2 < 0$.
From \eqref{eq: LGE} it is clear that the LGE eigenvalues scale trivially with $\mathcal{J}$.
Since we are only interested in the relative magnitude of the eigenvalues of the LGE, we can
without loss of generality set $J_1 = 1$ and for illustrative purposes $J_2 = -0.25$.
Let us choose a Fermi surface corresponding to $t_1=1$, $t_2=0$ and chemical potential $\mu=-2.5$.
Without SLP, the leading superconducting instability is $d_{xy}$-wave ($B_{2g}$ irrep).
The evolution of the LGE eigenvalues as we increase SLP is shown in \figref{fig: lieb main example}.
As SLP is increased, pairing between NN (different sublattices) is blocked, leading to suppression of the corresponding superconducting order. Strong SLP is still compatible with NNN (same sublattice) pairing. In this example, those bonds favor spin-triplet pairing, resulting in a leading $p$-wave ($E_u$ irrep) superconductivity when the SLP is strong. 

Let us investigate the properties of the resulting triplet superconducting order. For simplicity, we ignore the NN pairing by setting $J_1=0$, focusing only on the triplet states which can be formed on the NNN bonds with $J_2 = -0.25$.
In that case, we have four triplet bonds in the unit cell: ${\rm{AA}}_x$, ${\rm{AA}}_y$, ${\rm{BB}}_x$ and ${\rm{BB}}_y$ corresponding to pairing between the same sublattices in the $x$- and $y$-directions, respectively.
Importantly, in the presence of SLP these four bonds form two distinct sets that transform into themselves under a $C_4$ rotation (pairing across the plaquettes $\mathcal{S}_1$ and $\mathcal{S}_2$ respectively, see \figref{fig: illustration sublattice}). These sets are $\mathcal{S}_1 = \lbrace {\rm{AA}}_x, {\rm{BB}}_y\rbrace$ and $\mathcal{S}_2 =\lbrace {\rm{BB}}_x, {\rm{AA}}_y\rbrace$. 
The pairing magnitude of the two sets $\Delta_1$ and $\Delta_2$ will be different in the presence of SLP. Only in the absence of SLP, the two sets are related by symmetry (half-unit-cell translation).

\begin{figure}
    \centering
    \includegraphics[width=1.0\linewidth]{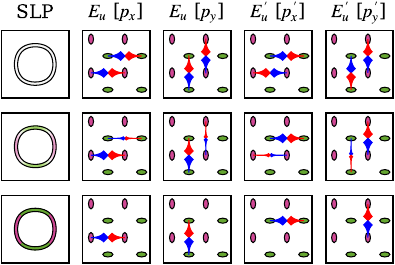}
    \caption{Real-space visualization of the longer-range triplet bonds as SLP increases. 
    The color and size of each bond is determined by the sign and magnitude of the superconducting pairing.
    In the absence of SLP, both bond sets $\mathcal{S}_1$ and $\mathcal{S}_2$ are related by translation symmetry. The $p$-wave states of the leading $E_u$ instability have the same sign on both bond sets, while for the subdominant $E_u'$ there is a sign change. As SLP increases, the two bond sets become inequivalent, resulting in a magnitude difference between the two sets. In the limit of extremely large SLP, the $p$-wave states are completely localized on one or the other set (here the leading instability is localized on $\mathcal{S}_2$).}
    \label{fig: triplet states real space}
\end{figure}

For concreteness, let us introduce notation to describe the superconducting states defined on these bonds.
The pairing amplitudes on all these four bonds are collected in a vector $\Vec{\Delta}$
\begin{equation} \label{eq: Delta set separation}
    \Vec{\Delta} = \Vec{\Delta}_1 \oplus \Vec{\Delta}_2,
\end{equation}
which can be separated using the two previously mentioned sets $\mathcal{S}_1$ and $\mathcal{S}_2$ with their corresponding 2D vectors
\begin{equation}
    \Vec{\Delta}_1 = 
    \begin{pmatrix}
        \Delta_{\rm{AA}}^x \\ \Delta_{\rm{BB}}^y 
    \end{pmatrix}, \quad
    \Vec{\Delta}_2 = 
    \begin{pmatrix}
        \Delta_{\rm{BB}}^x \\ \Delta_{\rm{AA}}^y 
    \end{pmatrix}.
\end{equation}
Considering triplet pairing, these four bonds will form two degenerate sets of eigenvalues, where each eigenspace has a $p_x$- and a $p_y$-component, regardless of SLP. In the absence of SLP, the eigenstates are
\begin{align}
    \Vec{\Delta}_x =  
    \begin{pmatrix}
        1 \\ 0
    \end{pmatrix}
    \oplus
    \begin{pmatrix}
        1 \\ 0
    \end{pmatrix}, \quad
    \Vec{\Delta}_y =  
    \begin{pmatrix}
        0 \\ 1
    \end{pmatrix}
    \oplus
    \begin{pmatrix}
        0 \\ 1
    \end{pmatrix}, \\
    \Vec{\Delta}'_x =  
    \begin{pmatrix}
        1 \\ 0
    \end{pmatrix}
    \oplus
    \begin{pmatrix}
        -1 \\ 0
    \end{pmatrix}, \quad
    \Vec{\Delta}'_y =  
    \begin{pmatrix}
        0 \\ 1
    \end{pmatrix}
    \oplus
    \begin{pmatrix}
        0 \\ -1
    \end{pmatrix},
\end{align}
where $\Vec{\Delta}_x$ and $\Vec{\Delta}_y$ are degenerate (similarly $\Vec{\Delta}'_x$ and $\Vec{\Delta}'_y$ are degenerate).
With only NNN pairing and without SLP, $\Vec{\Delta}_x$ and $\Vec{\Delta}_y$ are the leading instability while $\Vec{\Delta}'_x$ and $\Vec{\Delta}'_y$ have eigenvalue zero.
The reason for this is that without SLP, both sublattices are equivalent up to translation symmetry and having spatially modulating sign-alternating states is highly unfavorable.
These four states are shown in the first row of \figref{fig: triplet states real space}.

\begin{figure}
    \centering
    \includegraphics[width=1.0\linewidth]{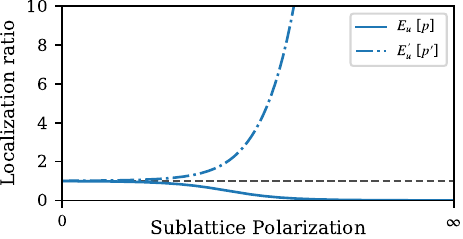}
    \caption{Localization ratio between the two bond sets $\mathcal{S}_1$ and $\mathcal{S}_2$ for the leading triplet states (see corresponding real space visualization of the states in \figref{fig: triplet states real space}) as SLP increases. For reference, the horizontal dashed line corresponds to equal magnitudes on both sets. As SLP increases, the triplet states become completely localized on one or the other set (here the leading $E_u$ instability becomes completely localized on $\mathcal{S}_2$). 
    }
    \label{fig: localization ratios triplet simple}
\end{figure}

Increasing SLP prevents hybridization between the two sublattices such that the $p_x$-state for each degenerate eigenvalue becomes more localized on one or the other set $\mathcal{S}_1$ or $\mathcal{S}_2$ (analogously for the $p_y$-state).
This is shown in the middle row of \figref{fig: triplet states real space}, where the $E_u$ states are primarily localized on $\mathcal{S}_2$ while the $E'_u$ states are primarily localized on $\mathcal{S}_1$.
In the limit of extreme SLP the eigenstates are completely localized on one or the other set.
In \figref{fig: localization ratios triplet simple}, we study the evolution of the localization ratio $|\Vec{\Delta}_1| / |\Vec{\Delta}_2|$ for each eigenvector as SLP increases.
Here we clearly see that as SLP increases, the $p$-wave states are localized mainly on one or the other set of bonds.

\begin{figure}
    \centering
    \includegraphics[width=1.0\linewidth]{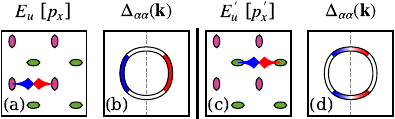}
    \caption{Real-space visualization and the Fermi surface gap structure for the $p_x$-components of the leading $E_u$ (a-b) and the subleading $E_u'$ (c-d) triplet instabilities, in the limit of extreme SLP.
    The leading instability takes advantage of the SLP structure, aligning the $p$-wave node (dashed line) with where there is no weight on the corresponding sublattice. In contrast, the subleading instability has the $p$-wave node where the weight on the corresponding sublattice is maximal.}
    \label{fig: triplet states band basis}
\end{figure}

The fact that the leading instability is mainly localized on $\mathcal{S}_2$ is not generic, but holds at least for sufficiently small Fermi surfaces which are not close to the BZ boundary.
The fact that $p$-wave states localized on $\mathcal{S}_2$ are favored in the limit of extreme SLP can be understood from the role of SLP in momentum space.
Let us focus on the $p_x$-state (the corresponding $p_y$-state behaves analogously). 
The first set $\mathcal{S}_1$ has a $p_x$-state localized on sublattice A, while $\mathcal{S}_2$ has a $p_x$-state localized on sublattice B.
The $\rm{BB}_x$ state is favorable, as it can take advantage of the momentum structure of the SLP by having its node already where superconductivity on sublattice B is suppressed. In contrast, $\rm{AA}_x$ introduces an additional node where superconductivity on sublattice A is maximal, which is energetically unfavorable. 
The clearest way to illustrate this difference is by looking at the superconducting order parameter in band space rather than sublattice space.
The gap matrix in sublattice space $\Delta^{(\rm{SL})}(\textbf{k})$ and in band space $\Delta^{(\rm{Band})}(\textbf{k})$ are related by the transformation
\begin{equation}
    \Delta^{(\rm{Band})}(\textbf{k}) = \unitary^\dagger (\textbf{k}) \Delta^{(\rm{SL})}(\textbf{k}) \unitary^* (-\textbf{k}),
\end{equation}
where $\unitary(\textbf{k})$ is the unitary transformation that diagonalizes the normal state Hamiltonian (i.e. $\mathcal{H}_0 (\textbf{k}) = \unitary (\textbf{k}) \Xi (\textbf{k}) \unitary^\dagger (\textbf{k})$, where $\Xi (\textbf{k}) = {\rm{diag}}(\xi_1 (\textbf{k}), \xi_2 (\textbf{k}))$).
We show the diagonal part of the gap matrix $\Delta_{\alpha \alpha} (\textbf{k})$ for the band $\alpha$ that crosses the Fermi surface in \figref{fig: triplet states band basis} for $p_x$-states for the leading and subleading triplet instabilities.
For the leading instability, the $p_x$-state is completely localized on sublattice B. The node at $k_x=0$ aligns with locations where there already is no sublattice weight on B. Considering that the corresponding instability for bonds on the A sublattice is a $p_y$ state with nodes at locations where there is no sublattice weight on A, one sees that the expected ground state has large gapping almost everywhere on the Fermi surface and two sign changes where the sublattice weight switches. This ``quasinodal'' state is driven by the SLP, but its density of states at low energies resembles quasinodal states discussed in relation to other multiband superconductors~\cite{Islam2025}, whereas in the latter case the very steep change of gapping is driven by extreme anisotropies in the interaction itself tuned close to a quantum critical point.
Contrast this with the $p_x$-state for the subleading instability.
It is completely localized on sublattice A, and the node at $k_x=0$ aligns precisely where the sublattice weight on A is maximal.

\subsubsection{Case 2: Leading singlet superconductivity} \label{sec: singlet}

In the previous example, we considered a case of exchange couplings $J_{\textbf{r}\textbf{r}'}$ which favored ``frustration-free'' N\'eel order (that is $J_1 > 0$ and $J_2 < 0$). 
This resulted in leading spin-triplet superconductivity supported by $J_2$ when NN pairing was suppressed by strong SLP.
However, it is not necessarily the case that the couplings $J_1$ and $J_2$ need to have different signs in order to support compensated magnetic order.
For example, compensated magnetic order could still be favored even if both $J_1$ and $J_2$ are positive, given that $J_2$ is significantly smaller than $J_1$.
In this section we explore the superconducting instabilities in such systems, still under the assumption that superconducting order develops rather than altermagnetism.
Similar to above, SLP prevents pairing on NN bonds, but will now lead to singlet superconductivity on the NNN bonds (since $J_2 > 0$). Here we discuss characteristic features of this longer-range singlet superconducting state.

\begin{figure}
    \centering
    \includegraphics[width=1.0\linewidth]{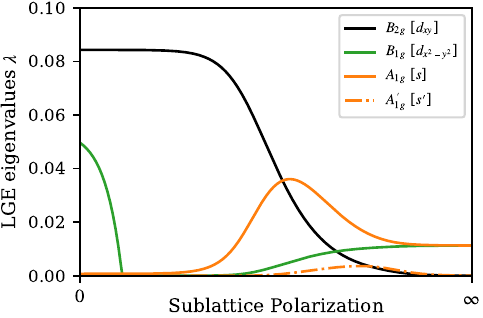}
    \caption{Eigenvalues of the linearized gap equation as SLP increases, where both the NN and NNN bonds favor singlet superconductivity ($J_1 =1$, $J_2 = 0.25$). As SLP increases, pairing on the NN bonds is suppressed. In the limit of strong SLP, the leading $s$- and $d$-wave singlet states are almost degenerate (i.e. have nearly the same critical temperature).}
    \label{fig: lieb main singlet both}
\end{figure}

Let us begin by considering the case $J_1=1$ and $J_2 = 0.25$, using the same Fermi surface as in the previous example.
Similar to before, we vary the SLP on a fixed Fermi surface and obtain the eigenvalues of the LGE shown in \figref{fig: lieb main singlet both}.
In the absence of SLP, the leading instability is $d_{xy}$-wave (that is $B_{2g}$) formed solely on the NN bonds.
The subdominant $d_{x^2 - y^2}$-wave state (that is $B_{1g}$) consists only of longer-range NNN bonds, while the $s$- and $s'$-states have weights on both NN and NNN bonds, in addition to onsite weights.
As SLP increases, pairing of different sublattices on the NN bond is suppressed, which we can see as the LGE eigenvalue associated with $d_{xy}$-wave vanishes. For the two $s$-wave states, the weights on the NN bonds vanish in strong SLP, resulting in leading $s$-wave states living entirely onsite and on the longer-range bonds. 
In the presence of SLP, onsite can also support $d_{x^2 - y^2}$-wave superconductivity, and the $B_{1g}$-channel therefore obtains onsite pairing weights. 

\begin{figure}
    \centering
    \includegraphics[width=1.0\linewidth]{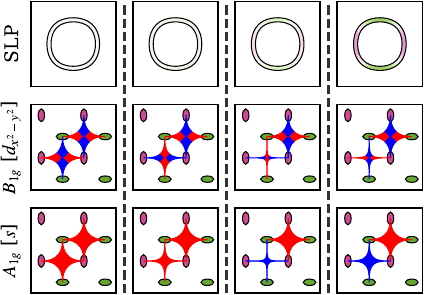}
    \caption{Visualization of the superconducting singlet state on NNN bonds as SLP increases.
    The first column shows the case without SLP.
    In the absence of SLP, $\Vec{\Delta}_1$ and $\Vec{\Delta}_2$ have the same sign for leading $B_{1g}$ and $A_{1g}$ instabilities (states $\Vec{\Delta}^{(0)}_d$ and $\Vec{\Delta}^{(0)}_s$ respectively). As SLP increases $\Vec{\Delta}^{(0)}_d$ and $\Vec{\Delta}^{(0)}_{d'}$ hybridize (likewise for $\Vec{\Delta}^{(0)}_s$ and $\Vec{\Delta}^{(0)}_{s'}$).
    This may induce a sign change between $\Vec{\Delta}_1$ and $\Vec{\Delta}_2$, indicated by the dashed lines.
    At the SLP for which the sign change occurs the superconducting state is completely localized on $\mathcal{S}_1$.
    Parameters used $J_1=0$, $J_2 = 0.25$.}
    \label{fig: real space singlet example d-wave}
\end{figure}

Interestingly, in the limit of strong SLP the leading singlet states form two pairs of almost degenerate $s$- and $d$-wave solutions.
In the following we investigate these states further.
For simplicity, we set $J_1=0$ and thereby ignore the NN bonds, so we can focus on properties of the leading singlet states.
Following the same procedure as in the triplet case, we have four longer-range singlet bonds, corresponding to pairing between the same sublattices in the $x$- and $y$-directions. 
We describe the longer-range superconducting bonds using the same state representation as \eqref{eq: Delta set separation}.
An important distinction is that the singlet states feature weights corresponding to onsite pairing. 
Let us here focus on the evolution of the longer-range bonds as we vary the SLP.
The symmetry of the superconducting singlet state is determined by the relative sign between the two bonds within each set ($\Delta^x_{\rm{AA}} = \pm \Delta_{\rm{BB}}^y$ and $\Delta_{\rm{AA}}^y = \pm \Delta_{\rm{BB}}^x$ for $s$- and $d$-wave, respectively).
In the absence of SLP, the four singlet eigenvectors of the linearized gap equation have the following NNN bonds
\begin{align}
    \Vec{\Delta}_s^{(0)}   = \begin{pmatrix}
        1 \\ 1
    \end{pmatrix} \oplus
    \begin{pmatrix}
        1 \\ 1
    \end{pmatrix}, \quad
     \Vec{\Delta}_d^{(0)}   = \begin{pmatrix}
        1 \\ -1
    \end{pmatrix} \oplus
    \begin{pmatrix}
        1 \\ -1
    \end{pmatrix}, \label{eq: A1g B1g no SLP} \\
    \Vec{\Delta}_{s'}^{(0)}   = \begin{pmatrix}
        1 \\ 1
    \end{pmatrix} \oplus
    \begin{pmatrix}
        -1 \\ -1
    \end{pmatrix},
    \quad
 \Vec{\Delta}_{d'}^{(0)}   = \begin{pmatrix}
        1 \\ -1
    \end{pmatrix} \oplus
    \begin{pmatrix}
        -1 \\ 1
    \end{pmatrix}.
\end{align}
In the example we are considering, $\Vec{\Delta}_d^{(0)}$ is the leading instability, while $\Vec{\Delta}_s^{(0)}$ is subleading.
The two remaining states, $\Vec{\Delta}_{s'}^{(0)}$ and $\Vec{\Delta}_{d'}^{(0)}$ both have eigenvalue 0 without SLP (similar to the triplet case).
The states $\Vec{\Delta}_d^{(0)}$ and $\Vec{\Delta}_s^{(0)}$ are depicted in the first column of \figref{fig: real space singlet example d-wave}.
As SLP increases, the two sublattices become distinct which results in a pairing difference between the two sets $\mathcal{S}_1$ and $\mathcal{S}_2$.
Let us first examine the limit of very large SLP since it will help us understand the behavior at finite SLP by interpolating between no and extreme SLP. At maximal SLP, the Fermi surface is completely sublattice polarized, such that the different sublattices A and B do not talk to each other. The LGE becomes block-diagonal, decoupling the two sublattice degrees of freedom. Importantly, the sublattices are related by rotational symmetry, which guarantees a twofold degeneracy of the same-sublattice pairing states. A natural choice for the two degenerate eigenstates is completely sublattice polarized superconducting states of the form
\begin{equation}
    \Vec{\Delta}_{\rm{A}} = 
    \begin{pmatrix}
        \alpha \\ 0
    \end{pmatrix}
    \oplus 
    \begin{pmatrix}
        0 \\ \beta
    \end{pmatrix}, \quad
    \Vec{\Delta}_{\rm{B}} = 
    \begin{pmatrix}
        0 \\ \alpha
    \end{pmatrix}
    \oplus 
    \begin{pmatrix}
        \beta \\ 0
    \end{pmatrix}.
\end{equation}
Note that for a specific sublattice, the weights on $x$ and $y$ are different, but rotational symmetry relates the two sublattices.
Using $\Vec{\Delta}_{\rm{A}}$ and $\Vec{\Delta}_{\rm{B}}$ we can construct another basis for this eigenspace whose basis vectors transform according to the irreducible representations (irreps) $A_{1g}$ and $B_{1g}$ respectively. These are
\begin{align}
        \Vec{\Delta}_s^{(\infty)} &  = \Vec{\Delta}_{\rm{A}} + \Vec{\Delta}_{\rm{B}} =
    \begin{pmatrix}
        \alpha \\ \alpha
    \end{pmatrix}
    \oplus 
    \begin{pmatrix}
        \beta \\ \beta
    \end{pmatrix} \label{eq: extreme SLP A1g}, \\
    \Vec{\Delta}_{d}^{(\infty)} & = \Vec{\Delta}_{\rm{A}} - \Vec{\Delta}_{\rm{B}} =
    \begin{pmatrix}
        \alpha \\ -\alpha
    \end{pmatrix}
    \oplus 
    \begin{pmatrix}
        -\beta \\ \beta
    \end{pmatrix}. \label{eq: extreme SLP B1g}
\end{align}
The fact that these two states are degenerate at maximal SLP, despite belonging to different irreps ($A_{1g}$ and $B_{1g}$, respectively) makes intuitive sense. What distinguishes these irreps is a sign-change under $C_4$ rotation. However, a $C_4$ rotation maps one sublattice into the other, and since the sublattices are completely decoupled, the relative sign between the sublattice-specific superconducting order parameters is irrelevant. Since $s$- and $d$-wave superconductivity are almost degenerate for strong SLP, this may lead to an instability towards $s+\ii d$ superconducting order which spontaneously breaks time-reversal symmetry at some temperature $T_{\rm{c}}^{\rm{TRSB}}$ below the superconducting critical temperature $T_{\rm{c}}$. 
The origin behind such an instability is that the complex linear combination would gap out nodes on the Fermi surface. 
Provided that SLP is sufficiently strong, this demonstrates that altermagnetic fluctuations may be more prone towards stabilizing $s + \ii d$ superconducting order in comparison to antiferromagnetic fluctuations.

\begin{figure}
    \centering
    \includegraphics[width=1.0\linewidth]{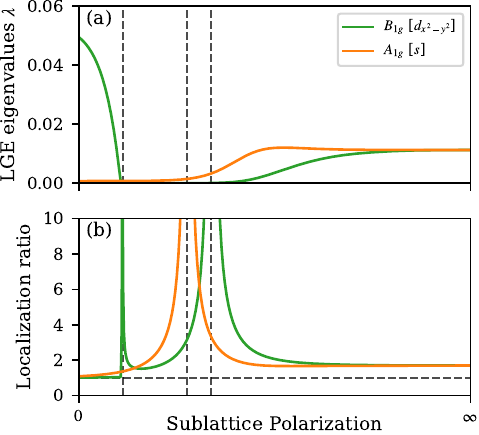}
    \caption{Eigenvalues of the linearized gap equation (a), and the localization ratio between the two sets $\mathcal{S}_1$ and $\mathcal{S}_2$ (b), as SLP increases.
    For some finite SLP (the vertical dashed lines, seen also in \figref{fig: real space singlet example d-wave}), the states are completely localized on $\mathcal{S}_1$.
    In the limit of strong SLP $s$- and $d$-wave are almost degenerate and mainly localized on $\mathcal{S}_1$.
    The horizontal dashed line in (b) corresponds to equal weight on $\mathcal{S}_1$ and $\mathcal{S}_2$.
    Parameters used $J_1=0$, $J_2 = 0.25$.}
    \label{fig: singlet localization ratios d-wave}
\end{figure}

Let us return to analyzing the superconducting states at extreme SLP, described by \eqref{eq: extreme SLP A1g} and \eqref{eq: extreme SLP B1g}. 
Compare these to the leading instabilities in the absence of SLP, described by \eqref{eq: A1g B1g no SLP}.
We will argue that as SLP increases, there must exist some finite SLP at which two of the eigenstates of the LGE are completely localized on either $\mathcal{S}_1$ or $\mathcal{S}_2$.
As SLP increases, the two leading instabilities must continuously go from $\Vec{\Delta}_s^{(0)}$ and $\Vec{\Delta}_d^{(0)}$ to $\Vec{\Delta}_{s}^{(\infty)}$ and $\Vec{\Delta}_{d}^{(\infty)}$ for each irrep, respectively.
Suppose that the eigenstate belonging to $A_{1g}$ goes from $\Vec{\Delta}_s^{(0)}$ to $\Vec{\Delta}_{s}^{(\infty)}$ without passing through $\Vec{\Delta}_1 = \Vec{0}$ or $\Vec{\Delta}_2 = \Vec{0}$. 
From \eqref{eq: A1g B1g no SLP} and \eqref{eq: extreme SLP A1g}, this requires $\alpha > 0$ and $\beta > 0$.
This implies that for the leading $B_{1g}$ channel, see \eqref{eq: A1g B1g no SLP} and \eqref{eq: extreme SLP B1g}, there must be a change in the sign between $\Vec{\Delta}_1$ and $\Vec{\Delta}_2$ at some finite SLP, meaning that either $\Vec{\Delta}_1$ or $\Vec{\Delta}_2$ needs to vanish at some finite intermediate SLP.
Had we assumed the opposite, i.e. that the leading $B_{1g}$ does not vanish on either $\mathcal{S}_1$ or $\mathcal{S}_2$, we would analogously arrive at the conclusion that $\Vec{\Delta}_1$ or $\Vec{\Delta}_2$ needs to vanish at some SLP for the leading $A_{1g}$ instability.
Looking at \figref{fig: real space singlet example d-wave}, this is precisely what happens. The last column shows the leading instabilities at large SLP. The $d$-wave channel has the same sign signature between the two sets as it did without SLP, while the leading $s$-wave channel now has picked up a sign difference between the two sets compared to the corresponding state without SLP.
This happens somewhere between the second and third column, where $s$- and $s'$-wave hybridize (as they belong to the same irrep).
In \figref{fig: singlet localization ratios d-wave}, we study the evolution of the LGE eigenvalues and the localization ratio $|\Vec{\Delta}_1| / |\Vec{\Delta}_2|$ for each eigenvector as SLP increases.
We see that this ratio diverges for the $s$-wave at some finite SLP.
Note that there are two values of SLP at which the localization ratio for the $d$-wave instability diverges (between first and second, and third and fourth column in \figref{fig: real space singlet example d-wave}). However, comparing the $d$-wave states in the absence of and in the extreme SLP limit, the relative sign between the two sets is the same.

\subsubsection{Experimental consequences and band space interpretation}

We have demonstrated that altermagnetic fluctuations result in a superconducting state with intra-unit-cell modulation.
In the case of leading spin-singlet superconductivity, the intra-unit-cell modulation can even be sign-changing.
In this section, we investigate the experimental consequences of this intra-unit-cell modulation.
Note that even in the nonmagnetic normal state of altermagnets, the translation symmetry relating the two sublattices is explicitly broken.
This is fundamentally different from an antiferromagnet, where the translation symmetry is spontaneously broken in the magnetic state and restored in the normal state.
In the case of antiferromagnetic fluctuations (that is the case without SLP), the modulated state (e.g. $\Vec{\Delta}_{s'}^{(0)}$) and the non-modulated state (e.g. $\Vec{\Delta}_{s}^{(0)}$) have clear differences in terms of experimental observables.
Let us first demonstrate this by studying the superconducting order parameter in band space rather than sublattice space.
Let us consider the same set of parameters as in \secref{sec: singlet}, where we for simplicity set $J_1=0$, such that the gap matrix is diagonal in sublattice space.
Even though the transformation to band space necessarily implies an off-diagonal gap matrix, the nodal structure of the quasiparticle dispersion is to zeroth order determined by the diagonal element for the band that crosses the Fermi surface. 
This is valid under the assumption that the gap matrix elements are significantly smaller than the electron dispersion for the band that does not cross the Fermi surface.

\begin{figure}
    \centering
    \includegraphics[width=1.0\linewidth]{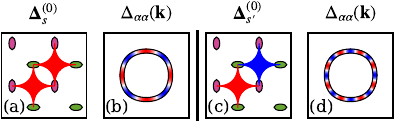}
    \caption{The real-space representation and the Fermi surface gap structure of the two $A_{1g}$ channels $\Vec{\Delta}_s^{(0)}$ and $\Vec{\Delta}_{s'}^{(0)}$ in the absence of SLP.
    The leading homogeneous $A_{1g}$ channel $\Vec{\Delta}_s^{(0)}$ is shown in (a) and (b), while the sign-changing $A_{1g}$ channel $\Vec{\Delta}_{s'}^{(0)}$ is shown in (c) and (d). The sign-changing state has additional nodes in the Fermi surface gap structure.}
    \label{fig: s-wave states without SLP}
\end{figure}

\begin{figure}
    \centering
    \includegraphics[width=1.0\linewidth]{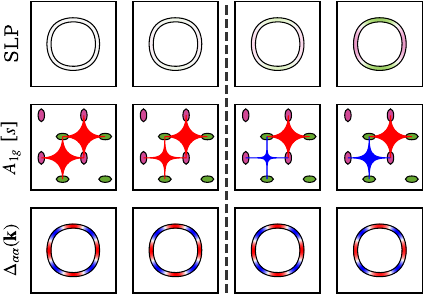}
    \caption{Evolution of the leading $s$-wave instability in real space and the gap structure on the Fermi surface as SLP increases. Despite the real space configuration acquiring a sign change (as indicated by the dashed line), the gap structure on the Fermi surface remains qualitatively the same.}
    \label{fig: s-wave leading instability momentum space}
\end{figure}

In \figref{fig: s-wave states without SLP} we show the real space and the diagonal element of the gap matrix in band space on the Fermi surface for the two $A_{1g}$ channels $\Vec{\Delta}_s^{(0)}$ and $\Vec{\Delta}_{s'}^{(0)}$ without SLP.
The sign-changing superconducting state $\Vec{\Delta}_{s'}^{(0)}$ spontaneously breaks translation invariance since $t_z=0$ and therefore formally can be considered as a pair-density wave (PDW) state. It is not a very competitive instability since it has additional nodes compared to the spatially homogeneous and leading $s$-wave state $\Vec{\Delta}_s^{(0)}$.
As we increase SLP, the translational symmetry is explicitly broken in the normal state Hamiltonian, and the two channels hybridize, resulting in the leading $A_{1g}$ channel shown in \figref{fig: real space singlet example d-wave}. From the perspective of the order parameter on the plaquettes, the state is still similar to the mentioned PDW state, however it is in this case forced to exist by the lower symmetry in the normal state already.
In \figref{fig: s-wave leading instability momentum space}, we show how the gap structure in band space evolves for the leading  $A_{1g}$ channel as SLP increases.
We see that even though the superconducting state in real space acquires a sign-alternating pattern as SLP increases, the gap structure on the Fermi surface remains essentially the same.
This illustrates that our intuition from \figref{fig: s-wave states without SLP} where the sign-alternating state behaves qualitatively differently does not apply in the presence of SLP.
A sign-alternating state in the presence of SLP can have qualitatively the same gap structure in band space as the homogeneous state in absence of SLP.

\begin{figure}
    \centering
    \includegraphics[width=1.0\linewidth]{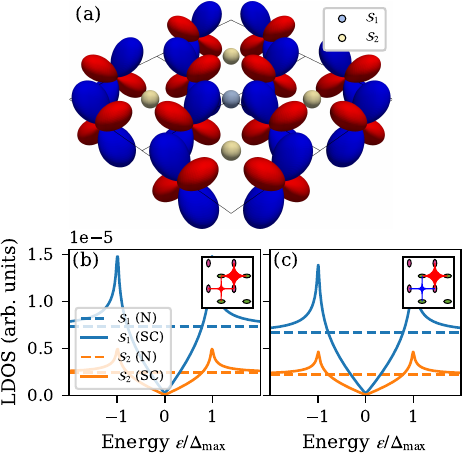}
    \caption{Panel (a) shows $d_{xz}$- and $d_{yz}$-orbitals located at the two respective sublattice degrees of freedom A and B of the Lieb lattice. 
    The blue (yellow) spheres indicate the location above which we expect a difference in the LDOS, originating from distinct overlap of the orbital wavefunctions.
    Panels (b-c) show the LDOS (both in the normal and superconducting state) measured above $\mathcal{S}_1$ and $\mathcal{S}_2$ for two distinct superconducting states at different values of SLP. 
    Both in the normal and superconducting state, the LDOS is larger at $\mathcal{S}_1$ from the asymmetric distribution of the orbital wavefunctions.
    Even though the second superconducting state is sign-alternating (see insets in (b) and (c)), no qualitative difference between the two superconducting states can be seen.
    }
    \label{fig: d-orbitals and LDOS}
\end{figure}

Let us discuss the potential consequences of the modulated superconducting state on the local density of states (LDOS) measured in scanning tunneling microscopy (STM).
In order to access the continuous LDOS (i.e. LDOS not restricted to the lattice sites) we assign orbital wavefunctions to the two sublattice degrees of freedom ($d_{xz}$ for sublattice A and $d_{yz}$ for sublattice B), see \figref{fig: d-orbitals and LDOS}~\cite{MerceLieb2026}.
More details on how the LDOS is calculated from the tight-binding model together with the orbital wavefunction is explained in \appendixref{appendix: LDOS}.
Since both the normal and superconducting state preserve $C_4$ rotational symmetry, the LDOS at the sublattice locations will be identical. 
A difference in the LDOS is however expected at the center of the two distinct plaquettes $\mathcal{S}_1$ and $\mathcal{S}_2$, when the two sublattice degrees of freedom are not related by translation symmetry (that is in the presence of $t_z$ and SLP).

In \figref{fig: d-orbitals and LDOS} we show the LDOS above $\mathcal{S}_1$ and $\mathcal{S}_2$ both in the normal state and the superconducting state for two values of SLP.
Importantly this translational symmetry is already explicitly broken in the normal state.
The LDOS is larger above $\mathcal{S}_1$ as a consequence of the distribution of the $d$-orbitals.
In the superconducting state, we observe a V-shaped gap opening at both locations, consistent with a nodal superconducting $s$-wave state, see \figref{fig: s-wave leading instability momentum space}.
As we increase SLP, we enter the sign-alternating state.
However, no qualitative difference compared to the non-sign-alternating superconducting state can be seen.

\subsubsection{Concrete example: $\rm{RuO}_2$}

Having demonstrated the potential consequences of SLP on the leading superconducting instability with two illustrative examples, we now turn to the concrete case of ${\rm{RuO}_2}$, one of the potential candidate materials for a ``failed altermagnet''~\cite{HiraishiRUO2,Kessler2024}. Interestingly, there is indeed evidence for induced superconductivity in strained RuO$_2$. While bulk RuO$_2$ shows no superconductivity down to $0.3$~K, films grown on TiO$_2$(110) substrates become superconducting at $T_{\rm{c}} \approx 2$~K~\cite{Uchida2020,Ruf2021}, and superconductivity has more recently also been realized in RuO$_2$(100) films~\cite{Wadehra2025}. The enhanced $T_{\rm{c}}$ has been attributed to a strain-induced redistribution of the Ru $d$-orbital occupancies that raises the density of states at the Fermi level, and electron--phonon estimates can account for its magnitude~\cite{Ruf2021}. The symmetry of the order parameter has, however, not been measured, and the present results suggest that the strong SLP of the RuO$_2$ bands could favor an unconventional triplet state if magnetic fluctuations contribute significantly to the pairing.

\begin{figure}
    \centering
    \includegraphics[width=1.0\linewidth]{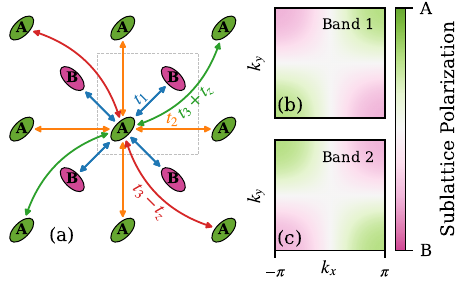}
    \caption{Illustration of the rutile lattice (a) and the sublattice polarization of the two bands (b-c). The dashed square in (a) outlines the unit cell.
    }
    \label{fig: rutile lattice}
\end{figure}

The normal state of RuO$_2$ has been previously modeled using a tight-binding formalism with two sublattice degrees of freedom (as in \eqref{eq: minimal model Hamiltonian}), where the hopping parameters have been chosen to capture important features of the band dispersions obtained from density functional theory (DFT)~\cite{Roig_Minimal_2024}.
An illustration of the lattice and the tight-binding hopping parameters are shown in \figref{fig: rutile lattice}(a).
Importantly, similar to the Lieb lattice, the two sublattice degrees of freedom are related by rotational symmetry and not translation symmetry, resulting in sublattice polarized bands.
The tight-binding parameters we use are $t_1=-0.425$, $t_2 = 0.05$, $t_3 = -0.025$, $t_z= 0.075$ and the chemical potential is set to $\mu = 0.2$~\cite{Roig_Minimal_2024}.
The parameter $t_z$, which controls the difference in third-nearest-neighbors hopping parameters, sets the degree of the SLP.
The SLP associated with the RuO$_2$ model is shown in \figref{fig: rutile lattice}(b-c).

\begin{figure}
    \centering
    \includegraphics[width=1.0\linewidth]{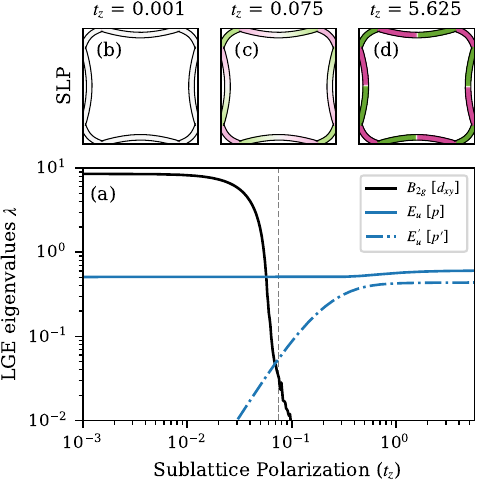}
    \caption{(a) Eigenvalues of the linearized gap equation for a model relevant for ${\rm{RuO}_2}$ versus increasing SLP. 
    The upper panels (b-d) show the degree of SLP on the Fermi surface for some selected values of $t_z$.
    For $t_z=0.075$ (indicated by the dashed line), the SLP is sufficiently strong to suppress the NN inter-sublattice singlet pairing, resulting in a leading triplet pairing state supported by the NNN bonds.
    }
    \label{fig: rutile with hubbard}
\end{figure}

As before, we vary the SLP of the bands by changing $t_z$ while keeping the Fermi surface fixed to that of ${\rm{RuO}_2}$ and calculate the eigenvalues of the leading superconducting instabilities of the Heisenberg interaction.
The exchange couplings in ${\rm{RuO}_2}$ have been estimated through DFT, with nearest-neighbor exchange $J_1=2.75$ and next-nearest-neighbor exchange $J_2=-1.8$~\cite{Smejkalmagnons}, both promoting ``frustration-free'' staggered magnetic order. 
For simplicity we ignore longer-range bonds.
The eigenvalues of the LGE scale trivially with $\mathcal{J}$ and therefore these values of exchange couplings are not connected to the units used in the tight-binding Hamiltonian.
In addition to the Heisenberg exchange coupling we include a strong Hubbard repulsion $U=100$.
The evolution of the LGE eigenvalues as we change the SLP is shown in \figref{fig: rutile with hubbard}.
We see that if the SLP structure of the bands is ignored, the leading instability is dominated by the NN inter-sublattice pairing bonds, resulting in a leading $B_{2g}$ instability.
As we increase SLP, the $B_{2g}$ eigenvalue vanishes, resulting in a leading triplet pairing state supported by the intra-sublattice NNN bonds.
The vertical dashed line indicates the value $t_z=0.075$, i.e. the value found to reproduce the DFT bands. For this value the SLP is sufficiently strong to suppress inter-sublattice pairing, resulting in a leading triplet instability.
This illustrates that the SLP of the bands could play a significant role in determining the leading superconducting instability in ${\rm{RuO}_2}$.

An important difference between the Lieb lattice and the rutile lattice is the role of onsite Hubbard repulsion on $B_{2g}$ pairing.
In the presence of SLP, onsite pairing couples to $B_{1g}$ in the Lieb lattice, and to $B_{2g}$ in the rutile lattice.
This means that as SLP increases, in the rutile lattice the $B_{2g}$ order stabilized by NN pairing couples to the onsite repulsion.
Consequently, suppression of the $B_{2g}$ eigenvalue is twofold: (i) SLP blocking of NN bonds (same as in the Lieb lattice), and (ii) coupling to the onsite repulsion.
Without Hubbard repulsion for $t_z=0.075$ corresponding to ${\rm{RuO}_2}$, $B_{2g}$ is still the leading instability with its eigenvalue suppressed to $60\%$ compared to the eigenvalue without SLP.
Contrast this with the case in \figref{fig: rutile with hubbard}, where the eigenvalue is only $3.5 \times 10^{-3}$ of its original value.
This demonstrates that for the rutile lattice, the coupling to the onsite Hubbard repulsion is the main reason for the suppression.
In contrast, for the Lieb lattice the suppression of the leading $B_{2g}$ instability is driven solely by the sublattice blocking mechanism.

\subsection{RPA altermagnetic-fluctuation pairing}
In the previous sections we demonstrated using the Heisenberg interaction in \eqref{eq: Heisenberg interaction} that the leading superconducting instability is highly affected by the SLP.
Here we show that this general trend can also be reproduced when interactions arise from a weak-coupling spin-fluctuation pairing mechanism.
To demonstrate this, we start from a minimal model description of the electronic structure on a square lattice~\cite{Roig_Minimal_2024} without altermagnetic crystal-field splitting $t_{z}$
and calculate the pairing interaction in momentum space within a two-sublattice spin-fluctuation approach~\cite{Romer2015} using the random phase approximation (RPA).
In this approach, we effectively allow pairing interactions of long-range nature. 
From the pairing we solve the analog of the linearized gap equation Eq.~(\ref{eq: LGE}), but where the vector of order parameters now is parametrized by momenta on the Fermi surface~\cite{Graser2009,Kreisel13}.
For more details, see \appendixref{appendix: RPA pairing}.

\begin{figure}
    \centering
    \includegraphics[width=1.0\linewidth]{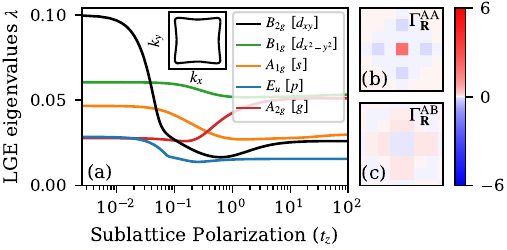}
    \caption{Leading superconducting instabilities induced by altermagnetic fluctuations from a weak-interaction RPA approach. 
    Panel (a) shows the evolution of the leading eigenvalues of the linearized gap equation as SLP is increased. 
    The inset shows the Fermi surface.
    Panels (b) and (c) show the effective pairing between same and different sublattices, respectively.
    As SLP increases, the symmetry of the leading instability changes, consistent with the previous description using the Heisenberg model.
    } \label{fig: RPA case 1}
\end{figure}

\begin{figure}
    \centering
    \includegraphics[width=1.0\linewidth]{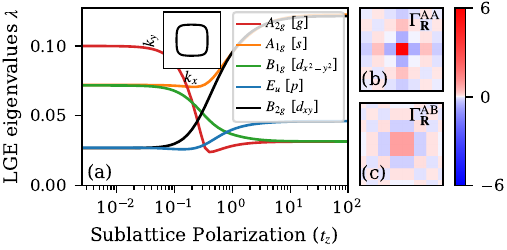}
    \caption{Same as in \figref{fig: RPA case 1} but for a different Fermi surface.
    } 
    \label{fig: RPA case 2}
\end{figure}

We consider a square lattice with $t_1 = 0.5$ and $t_2=-0.15$, and calculate the effective pairing interaction for two Fermi surfaces by setting the chemical potential to $\mu_1 = -0.7$ and $\mu_2 = -1.05$, shown in \figref{fig: RPA case 1} and \figref{fig: RPA case 2} respectively.
The weak-coupling spin-fluctuation calculation produces a number of attractive pairing interactions appearing as blue squares in panels (b, c)~\cite{Kreisel2015,Choubey2014}. 
Panel (b) shows the interaction between same sublattices $\Gamma_{\textbf{R}}^{\rm{AA}}$ while panel (c) shows the interaction between different sublattices $\Gamma_{\textbf{R}}^{\rm{AB}}$.
Note that the center of each panel refers to $\textbf{R}=\textbf{0}$ and neighboring pixels correspond to neighboring unit cells.
The onsite Hubbard repulsions are chosen as $U_1=0.85$ and $U_2 = 1.68$ such that the leading instability has the same eigenvalue $\lambda =0.1$ in both cases.
For the case of a slightly nested Fermi surface in \figref{fig: RPA case 1}, it turns out that the nearest-neighbor interaction (center of panel (c)) is attractive, resulting in a leading $B_{2g}$ superconducting instability. 
For the second choice of chemical potential in \figref{fig: RPA case 2}, the Fermi surface is less nested (inset of panel (a)), the spin susceptibility exhibits a broader structure~\cite{Romer2015,RomerRMP} and the pairing interactions have a more long-range structure, introducing frustration in many pairing channels, resulting in leading $A_{2g}$ ($g$-wave) instability. This frustration is consistent with the fact that the effective pairing in the second case needs to be noticeably larger to produce the same largest eigenvalue $\lambda = 0.1$ as in the first case.

We now examine the effect of SLP in the electronic structure by tuning $t_{z}$ (as in the rutile lattice, see Fig.~\ref{fig: rutile lattice}(a)), while keeping the pairing interaction and the Fermi surface unchanged.
As shown in \figref{fig: RPA case 1}(a), the $B_{2g}$ instability quickly becomes less competitive, leaving the $B_{1g}$ instability with the largest eigenvalue since it can take advantage of the longer-range intra-sublattice pairing. 
For the second case shown in \figref{fig: RPA case 2}(a), SLP decreases the eigenvalue of the leading $A_{2g}$ instability while allowing for an increase in the eigenvalue for other channels such that eventually $A_{1g}$ becomes dominant because the frustrating inter-sublattice interactions are effectively removed by the sublattice polarized bands.
For both Fermi surfaces, we note that in the limit of strong SLP, we have two pairs of eigenvalues that are almost degenerate: ($A_{1g}$, $B_{2g}$), and ($B_{1g}$, $A_{2g}$).
The mechanism behind this is precisely what was described in \secref{sec: singlet}, where we identified almost degenerate $A_{1g}$ and $B_{1g}$ states in the case of the Lieb lattice.
Which $d$-wave state is degenerate with the $s$-wave state depends on the structure of the SLP associated with the symmetry of the crystal ($B_{1g}$ for Lieb and $B_{2g}$ for rutile, see \figref{fig: illustration sublattice} and \figref{fig: rutile lattice} respectively).
$B_{1g}$ and $A_{2g}$ are almost degenerate in the rutile lattice for the same reason. 
What distinguishes these two states is the additional nodes for $A_{2g}$ at $k_x=0$ and $k_y=0$, which align with the nodes of the SLP.
For extremely large SLP, the sublattices are almost decoupled, and the relative sign between the two sublattice degrees of freedom is irrelevant.

\section{Conclusions}

In this study we have investigated how altermagnetic fluctuations impact the superconducting pairing instabilities.
We have focused on the role played by the so-called sublattice polarization (SLP), a feature of the nonmagnetic state characteristic of altermagnets (in contrast to simple ferromagnets and antiferromagnets).
By mean-field decoupling of a Heisenberg interaction Hamiltonian \eqref{eq: Heisenberg interaction} in the Cooper channel, we demonstrated that SLP blocks nearest-neighbor (NN) pairing, which typically is the dominant unconventional pairing mechanism.
SLP suppresses the NN pairing, resulting in a longer-range same-sublattice pairing state which can differ significantly from the NN dominated pairing. 
In altermagnets, the NN bonds favor staggered magnetism and consequently singlet superconductivity.
To obtain staggered magnetism without frustrated exchange coupling, the longer-range next-nearest-neighbor (NNN) bonds in contrast favor parallel spins and consequently triplet pairing.
It is important to stress that this triplet state is fundamentally different from the equal-spin pairing triplet state proposed to coexist with magnetic order~\cite{Zhu2023,Hong2025,Monkman,Hestbjerg,Zou2026}.
Weak NNN bonds that favor anti-parallel spins are still compatible with staggered magnetic order.
In this case, the leading longer-range pairing state induced by SLP is singlet, yet differs from the NN singlet state.
We demonstrated that SLP may induce a sign-changing intra-unit-cell modulated state.
It is important to stress that the translational symmetry is explicitly broken in the normal state (since in the presence of SLP, the two sublattices are not related by translation symmetry).
We discussed that as the sign-changing  PDW state is entered, there is no qualitative change in the superconducting order parameter in band space, or in observables such as local density of states.
However, in the limit of extremely strong SLP, the two leading NNN $A_{1g}$ ($s$-wave) and $B_{1g}$ ($d$-wave) instabilities become almost degenerate, demonstrating the possibility of forming an $s + \ii d$ state and spontaneously breaking time-reversal symmetry in addition to ${\rm{U}}(1)$ symmetry.
Lastly, we investigated a Hamiltonian where the effective pairing interaction was obtained from dressing the bare Hubbard repulsion through the RPA pairing diagrams.
We demonstrated that, similar to the case of the Heisenberg Hamiltonian, increasing SLP can qualitatively change the leading superconducting instability.

\begin{acknowledgments}
We acknowledge useful discussions with D. F. Agterberg and S. C. Holb\ae k. M.B. was supported by a research grant (VIL69220) from VILLUM FONDEN. B.M.A. acknowledges support from the Independent Research Fund Denmark Grant No. 5241-00007B. A.K. acknowledges support by the Danish National Committee for Research Infrastructure (NUFI) through the ESS-Lighthouse Q-MAT.
\end{acknowledgments}

\appendix

\section{Derivation of the linearized gap equation} \label{appendix: LGE derivation}

In this section we derive the linearized gap equation \eqref{eq: LGE} used to analyze the superconducting instabilities arising from the Heisenberg interaction \eqref{eq: Heisenberg interaction}.
Note that the Hubbard repulsion can be included as an onsite Heisenberg term, and therefore it is sufficient to include the generalized form of the Heisenberg interaction.
Using $S_i (\textbf{r}) =  \Vec{c}^\dagger (\textbf{r}) \sigma_i \Vec{c} (\textbf{r})$, where $\Vec{c}^\dagger = \begin{pmatrix}
    c_\uparrow^\dagger & c_\downarrow^\dagger
\end{pmatrix}$, we obtain
\begin{align*}
    H_\inter  =  -\sum_{\textbf{r}\textbf{r}'}J_{\textbf{r}\textbf{r}'} & \Big( c_\uparrow^\dagger (\textbf{r}) c_\uparrow (\textbf{r}) c_\downarrow^\dagger (\textbf{r}') c_\downarrow (\textbf{r}') + 
    c_\downarrow^\dagger (\textbf{r}) c_\downarrow (\textbf{r}) c_\uparrow^\dagger (\textbf{r}') c_\uparrow (\textbf{r}')  \\
    & + 2[c_\uparrow ^\dagger (\textbf{r}) c_\uparrow (\textbf{r}') c_\downarrow ^\dagger (\textbf{r}') c_\downarrow (\textbf{r}) + \rm{H.c.} ] \Big),
\end{align*}
where we have kept only the terms leading to opposite spin pairing (once decoupled in the Cooper channel). 
This restriction can be made without loss of generality since both our interaction and normal state Hamiltonian are invariant under spin rotations. 
Assuming translation invariance (i.e. the exchange couplings depend on the difference $\textbf{r}- \textbf{r}'$), it is convenient to replace the sum over bonds by a sum over all positions and separations. In addition we introduce sublattice degrees of freedom $\textbf{r} = (\textbf{R}, a)$ such that the interaction reads
\begin{align*}
    H_\inter  =  -\sum_{\textbf{R},\Vec{\delta}}
    \sum_{a,b}
    & J_{\Vec{\delta}}^{ab}  \Big( c_{\uparrow a}^\dagger (\textbf{R}) c_{\uparrow a} (\textbf{R}) c_{\downarrow b}^\dagger (\textbf{R} + \Vec{\delta}) c_{\downarrow b} (\textbf{R} + \Vec{\delta}) \\
    & + [c_{\uparrow a} ^\dagger (\textbf{R}) c_{\uparrow b} (\textbf{R} + \Vec{\delta}) c_{\downarrow b} ^\dagger (\textbf{R} + \Vec{\delta}) c_{\downarrow a} (\textbf{R}) + \rm{H.c.} ] \Big).
\end{align*}
By going to reciprocal space
$$
c_{\sigma a}( \textbf{R}) = \int_{\textbf{k} \in \BZ} \frac{\dd ^2 \textbf{k}}{|\Omega_\BZ|}  \ee^{\ii \textbf{k} \cdot \textbf{R} }c_{\sigma a}(\textbf{k}),
$$
and
$$
J^{ab} (\textbf{k}) = \sum_{\Vec{\delta}}
J^{ab}_{\Vec{\delta}} \ee ^{-\ii \textbf{k} \cdot \Vec{\delta}},
$$
we obtain, after mean-field decoupling in the Cooper channel 
\begin{equation}
    H_{\inter}^{(\MF)} = \int \frac{\dd ^2 \textbf{k}}{|\Omega_\BZ|} \Delta_{ab}(\textbf{k}) c^\dagger_{\uparrow a} (\textbf{k}) c^\dagger _{\downarrow b}(-\textbf{k}) + \rm{H.c.},
\end{equation}
with
\begin{equation}
    \begin{aligned}
        \Delta_{ab}(\textbf{k}) = \int  & \frac{\dd ^2 \textbf{k}'}{|\Omega_\BZ|}
    \Big(
     J^{ab}(-\textbf{k}+\textbf{k}') \expval{c_{\uparrow a} (\textbf{k}') c_{\downarrow b} (-\textbf{k}')}
 \\
    &+  2J^{ab}(-\textbf{k}-\textbf{k}') \expval{c_{\uparrow b} (\textbf{k}') c_{\downarrow a} (-\textbf{k}')}
    \Big).
    \end{aligned} \label{eq: Appendix self-consistent k-space}
\end{equation}
The linearized gap equation is obtained by expanding the anomalous expectation values to first order in $\Delta_{ab}(\textbf{k})$, treating $H_\inter^{(\MF)}$ perturbatively on top of the normal state Hamiltonian. This gives 
\begin{equation}
    \expval{c_{\uparrow a} (\textbf{k})  c_{\downarrow b} (-\textbf{k})} =
    \sum_{cd }W_{abcd}(\textbf{k}) \Delta_{cd}(\textbf{k}) + \mathcal{O}(\Delta^2),
\end{equation}
where
\begin{equation}
    \begin{aligned}
        W_{abcd}(\textbf{k}) = \sum_{\alpha \beta} &  \unitary_{a \alpha} (\textbf{k})\unitary_{b \beta}(-\textbf{k}) \unitary_{c \alpha}^* (\textbf{k}) \unitary_{d \beta}^* (-\textbf{k})  \\
        & \times\frac{1-n_\FD [\xi_\alpha (\textbf{k})] - n_\FD [\xi_\beta (-\textbf{k})]}{\xi_\alpha (\textbf{k}) + \xi_\beta (-\textbf{k})}.
    \end{aligned}
\end{equation}
Here $\alpha$ and $\beta$ are band indices and $\unitary$ is the unitary transformation that diagonalizes the normal state Hamiltonian — that is $\mathcal{H}_0 = \unitary \Xi \unitary^\dagger$ with $\Xi = {\rm{diag}}(\xi_1, \xi_2, \ldots)$ being the normal state band dispersions, and $n_\FD (\epsilon ) = (1 + \exp(\epsilon / k_{\rm{B}}T))^{-1}$ is the Fermi-Dirac distribution.
Inserting this into the self-consistency equation \eqref{eq: Appendix self-consistent k-space} we obtain the linearized gap equation written in reciprocal space as
\begin{equation}
    \begin{aligned}
        \Delta_{ab}(\textbf{k}) = \sum_{cd}
        \int & \frac{\dd ^2 \textbf{k}'}{|\Omega_\BZ|} 
    \Big( 
    J^{ab}(-\textbf{k} + \textbf{k}')W_{abcd} (\textbf{k}')
    \\
    +
    & 2J^{ab}(-\textbf{k} - \textbf{k}')W_{bacd} (\textbf{k}')
    \Big) \Delta_{cd}(\textbf{k}').
    \end{aligned} \label{eq: Appendix LGE k-space}
\end{equation}
Since we are dealing with only pairing on a finite set of bonds, it is convenient to parametrize the superconducting order parameter in terms of these bonds.
Note that all dependence on $\textbf{k}$ is embedded in
$$
J^{ab}(-\textbf{k} \pm \textbf{k}') = \sum_{\Vec{\delta}} J_{\Vec{\delta}} ^{ab} \eta_{\Vec{\delta}} (\textbf{k})\eta_{\Vec{\delta}}^*(\pm \textbf{k}'),
$$
where we have introduced the form factors $\eta_{\Vec{\delta}} (\textbf{k}) = \ee^{i \textbf{k}\cdot \Vec{\delta}}$.
This allows us to write the superconducting order parameter as
\begin{equation}
    \Delta_{ab}(\textbf{k}) = \sum_{\Vec{\delta}} \eta_{\Vec{\delta}}(\textbf{k})\Delta_{ab}^{\Vec{\delta}},
\end{equation}
where $\Delta_{ab}^{\Vec{\delta}}$ are some weights (associated with the form factors for each bond). The linearized gap equation can be written in terms of these weights as
\begin{equation}
    \begin{aligned}
        \Delta_{ab}^{\Vec{\delta}} = J_{\Vec{\delta}}^{ab} \sum_{cd, \Vec{\delta '}}\int  
        &\frac{\dd ^2 \textbf{k}}{|\Omega_\BZ|}
    \Big(
    \eta_{\Vec{\delta}}^* (\textbf{k})W_{abcd} (\textbf{k}) \\
    &+ 
    2\eta_{\Vec{\delta}}^* (-\textbf{k})W_{bacd} (\textbf{k})
    \Big) \eta_{\Vec{\delta'}}(\textbf{k}) \Delta_{cd}^{\Vec{\delta'}}.
    \end{aligned}
\end{equation}
For practical reasons, dealing with this version of the linearized gap equation is computationally more tractable, since it involves diagonalizing a matrix of size $N_{\rm{bonds}} \times N_{\rm{bonds}}$, where $N_{\rm{bonds}}$ denotes the number of pairing bonds, which is significantly less (for the Heisenberg model we are using) than the number of points in reciprocal space.

Lastly we go to the BCS limit — that is for weak interaction strengths such that $k_{\rm{B}}T_{\rm{c}} \ll \hbar \omega_{\rm{D}} \ll t$, where $ \hbar \omega_{\rm{D}}$ is the energy cut-off for pairing, and $t$ is the characteristic energy scale of the normal state Hamiltonian (e.g. hopping parameter).
The inequality $\hbar \omega_{\rm{D}} \ll t$ allows us to restrict the integration to points close to the Fermi surface and importantly separate the integration along and perpendicular to the Fermi surface. 
The variation in the form factors and the unitary transformation perpendicular to the Fermi surface are small, such that these can be moved completely inside the integral along the Fermi surface.
In $W_{abcd}(\textbf{k})$, only intraband terms ($\alpha=\beta$) for bands that cross the Fermi surface give the dominant contribution.
Change of variables converts the integral perpendicular to the Fermi surface into an integral over energy $\xi$, leaving us with the standard BCS integral
\begin{equation}
    \int_0 ^{\hbar \omega_{\rm{D}}} \frac{\tanh ( \xi / 2k_{\rm{B}}T_{\rm{c}}) }{\xi} \dd{\xi}
     \approx
     \ln \left( 
     1.13
     \frac{\hbar \omega_{\rm{D}}}{k_{\rm{B}}T_{\rm{c}}}
     \right),
\end{equation}
where we assumed that the bands are inversion symmetric $\Xi (-\textbf{k}) = \Xi (\textbf{k})$ and used the inequality $k_{\rm{B}}T_{\rm{c}} \ll \hbar \omega_{\rm{D}}$.
We can now express the matrix elements of the linearized gap equation as integrals along the Fermi surface.
To shorten notation, let us introduce a generalized index $X=(ab, \Vec{\delta})$ and $X'=(a'b', \Vec{\delta'})$. 
We can write the linearized gap equation in terms of these weights as
\begin{equation}
    \Delta_X =  \ln \left( 
     1.13
     \frac{\hbar \omega_{\rm{D}}}{k_{\rm{B}}T_{\rm{c}}}
     \right) J_X \sum_{X'} \mathcal{M}_{XX'} \Delta_{X'} ,
\end{equation}
where
\begin{equation} \label{eq: LGE M-matrix}
    \begin{aligned}
        & \mathcal{M}_{XX'} =  \sum_{\alpha}\frac{1}{|\Omega_\BZ|}\oint_{\rm{FS}^{(\alpha)}} \frac{\dd  \ell}{|\Vec{v}_{\rm{F}}^{(\alpha)}|}
    \Big( 
    \unitary_{a \alpha} (\textbf{k})\unitary_{b \alpha}(-\textbf{k}) \eta_{\Vec{\delta}}^* (\textbf{k})
    \\
    & \quad +
    2\unitary_{b \alpha} (\textbf{k})\unitary_{a \alpha}(-\textbf{k}) \eta_{\Vec{\delta}}^* (-\textbf{k})
    \Big) 
    \unitary_{a' \alpha}^* (\textbf{k}) \unitary_{b' \alpha}^* (-\textbf{k}) \eta_{\Vec{\delta'}}(\textbf{k}),
    \end{aligned}
\end{equation}
where we sum over the bands $\alpha$ that cross the Fermi level and $\Vec{v}_{\rm{F}}^{(\alpha)} = \Vec{\nabla} \xi_\alpha$ is the associated Fermi velocity.
From this we can identify the linearized gap equation $\lambda \Vec{\Delta} = \mathcal{JM} \Vec{\Delta}$ (that is \eqref{eq: LGE} from the main text), where we can relate the largest eigenvalue $\lambda_{\max}$ to the critical temperature as
\begin{equation}
k_{\rm{B}}T_{\rm{c}} \approx 1.13 \hbar \omega_{\rm{D}} \ee ^{-1/\lambda_{\max}}.
\end{equation}

\begin{figure}
    \centering
    \includegraphics[width=0.7\linewidth]{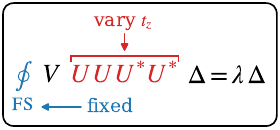}
    \caption{
    Schematic of the LGE for some general pairing interaction $V$. 
    The Fermi surface is calculated for some fixed Hamiltonian.
    The sublattice polarization is varied through the unitary transformation $U (\textbf{k})$, which is calculated as $t_z$ changes.
    }
    \label{fig: LGE sketch}
\end{figure}

Lastly, let us further elaborate on how we solve the linearized gap equation as we vary the sublattice polarization, while keeping the Fermi surface fixed.
The LGE has three ingredients: (i) an interaction $V$, (ii) the Fermi surface, and the associated Fermi velocity, and (iii) the unitary transformation $\unitary (\textbf{k})$, see \eqref{eq: LGE M-matrix}.
The Fermi surface and Fermi velocity are obtained from a starting Hamiltonian $\mathcal{H}_0 (\textbf{k})$.
The unitary transformation is obtained from another Hamiltonian $\tilde{{\mathcal{H}}}_0 (\textbf{k})$.
We vary $\tilde{t}_z$ which is the parameter that sets the SLP of the eigenstates associated with $\tilde{{\mathcal{H}}}_0 (\textbf{k})$, which in turn alters the unitary matrix $\unitary (\textbf{k})$.
An equivalent way to formulate this procedure is to say that we consider Hamiltonians $\mathcal{H}_0 (\textbf{k}) = \unitary (\textbf{k}) \Xi (\textbf{k}) U^\dagger (\textbf{k}) $, where we keep the band dispersions contained in $\Xi (\textbf{k})$ fixed, while we vary the unitary matrix $U (\textbf{k})$ through $t_z$.
This allows us to study the LGE eigenvalues and eigenvectors dependence on SLP, isolating the key feature that qualitatively distinguishes altermagnetic fluctuations from ferromagnetic and antiferromagnetism fluctuations.
This methodology is illustrated in \figref{fig: LGE sketch}.

\section{Continuous Local Density of States} \label{appendix: LDOS}

In this section we explain the details on how the LDOS calculations in the main text were carried out.
The local density of states at position $\textbf{r}$ and energy $\epsilon$ is defined as
\begin{equation}
    \rho (\textbf{r}, \epsilon) = \sum_{n} |\psi_n (\textbf{r})|^2 \delta (\epsilon -\epsilon_n),
\end{equation}
where $\psi_n (\textbf{r})$ is the quantum eigenstate with eigenenergy $\epsilon_n$. In the tight-binding formalism, we can model $\psi_n (\textbf{r})$ using the tight-binding eigenstate $u_n (\textbf{r})$, where $\textbf{r}$ denotes the discrete positions of the lattice.
Specifically, we set
\begin{equation}
    \psi_n (\textbf{r}) =  \sum_{\textbf{r}'} u_n (\textbf{r}') \phi_{\textbf{r}'}(\textbf{r} - \textbf{r}'),
\end{equation}
where $\phi_{\textbf{r}'}(\textbf{r} - \textbf{r}')$ is the orbital wavefunction associated with site $\textbf{r}'$ in the lattice and $u_n (\textbf{r}')$ is the tight-binding eigenstate.
Let us first discuss how to calculate the LDOS in the normal state (the calculation in the superconducting state follows analogously).
It is convenient to write the eigenstates using Bloch states, which are obtained by diagonalizing the corresponding Bloch Hamiltonian in reciprocal space $\textbf{k}$.
We introduce sublattice degrees of freedom ($a$) and band degrees of freedom ($\alpha$) and write
\begin{equation}
    \rho (\textbf{r}, \epsilon) = \sum_\alpha \int \frac{\dd ^2 \textbf{k}}{|\Omega_\BZ|} |\psi_{\alpha, \textbf{k}} (\textbf{r})|^2 \delta (\epsilon - \xi_{\alpha} (\textbf{k})),
\end{equation}
where
\begin{equation} \label{eq: psi for LDOS}
    \psi_{\alpha, \textbf{k}} (\textbf{r}) = \sum_{a} \unitary_{a \alpha} (\textbf{k}) \chi_{a, \textbf{k}} (\textbf{r}),
\end{equation}
and
\begin{equation} \label{eq: chi for LDOS}
    \chi_{a, \textbf{k}} (\textbf{r}) = \sum_{\textbf{R}} \ee^{\ii \textbf{k} \cdot \textbf{R}} \phi_{a}(\textbf{r}-\textbf{r}_{\textbf{R},a}),
\end{equation}
where $\phi_a$ is the orbital wavefunction associated with sublattice $a$ and $\textbf{r}_{\textbf{R},a}$ is the position of the atom located in unit cell $\textbf{R}$ at sublattice $a$. 
For a given real space position $\textbf{r}$ it is often sufficient to restrict the sum over unit cells in \eqref{eq: chi for LDOS} to unit cells close to the desired real space positions (since the contribution from the remaining orbital wavefunctions is negligible).
The superconducting state is described by the Nambu matrix
\begin{equation}
    \mathcal{H}_{\rm{Nambu}} (\textbf{k}) =
    \begin{pmatrix}
        \mathcal{H}_0 (\textbf{k}) & \Delta (\textbf{k}) \\
        \Delta^\dagger (\textbf{k}) & -\mathcal{H}_0^{\rm{T}} (-\textbf{k})
    \end{pmatrix},
\end{equation}
with eigenvectors $\Vec{w}_\alpha (\textbf{k})$ and quasiparticle eigenenergies $E_\alpha (\textbf{k})$  expressed as
\begin{equation}
    \mathcal{H}_{\rm{Nambu}}(\textbf{k}) \Vec{w}_\alpha (\textbf{k}) =  E_\alpha (\textbf{k}) \Vec{w}_\alpha (\textbf{k}).
\end{equation}
The eigenvector can be partitioned into an electron-like and hole-like part
\begin{equation}
    \Vec{w}_\alpha (\textbf{k}) = \begin{pmatrix}
        \Vec{u}_{\alpha} (\textbf{k}) \\
        \Vec{v}_{\alpha} (\textbf{k})
    \end{pmatrix}.
\end{equation}
To calculate the LDOS in the superconducting state, we can use precisely the same formalism as in the normal state, but replace the band dispersions $\xi_\alpha (\textbf{k})$ with the quasiparticle energies $E_{\alpha} (\textbf{k})$ and the unitary transformation $\unitary_{a \alpha} (\textbf{k})$ with the electron-like part $u_{a \alpha} (\textbf{k})$ (note that the sum over $\alpha$ contains twice as many terms).

In the BCS limit, we can express the LDOS in terms of quasiclassical theory for superconductivity. At the energy scale of the superconducting gap, the LDOS can be expressed as an integral over the Fermi surface
\begin{equation}
                \rho (\textbf{r}, \epsilon) = \sum_{\alpha}\frac{1}{|\Omega_\BZ|}
            \oint_{\rm{FS}^{(\alpha)}}
            \frac{\dd  \ell}{|\Vec{v}_{\rm{F}}^{(\alpha)}|}
            |\psi_{\alpha, \textbf{k}} (\textbf{r})|^2 |g_\alpha(\textbf{k}, \epsilon)|,
\end{equation}
where
\begin{equation}
    g_\alpha(\textbf{k}, \epsilon) =
     \lim_{\eta \to 0^+} {\rm{Re}} \left[ 
            \frac{\epsilon + \ii \eta}{\sqrt{(\epsilon + \ii \eta)^2 - |\Delta_{\alpha \alpha} (\textbf{k})|^2}}
            \right].
\end{equation}
Here $\Delta_{\alpha \alpha} (\textbf{k})$ is the intraband component of the pairing matrix in band space for the band that crosses the Fermi surface.

Having explained the general methodology, let us now specify what orbital wavefunctions we use in the case of the Lieb lattice, and at which positions (in three-dimensional space) we calculate the LDOS.
We calculate the LDOS at $z=a_0$, where $a_0$ is the distance between same sublattice degrees of freedom (that is the unit cell is of size $a_0 \times a_0$ and the nearest neighbor distance is $a_0 / \sqrt{2}$).
The $d_{xz}$-orbitals on sublattice A are modeled as
\begin{equation}
    \phi_{\rm{A}} (x,y,z) = xz \cdot  \ee^{-\sqrt{x^2+y^2+z^2}/\ell_0},
\end{equation}
where $\ell_0 = 0.2 a_0$.
The $d_{yz}$-orbitals on sublattice B are analogously defined by interchanging $x \leftrightarrow y$.
\section{Weak coupling spin-fluctuation pairing calculations} \label{appendix: RPA pairing}

In real materials, the pairing interaction includes contributions from further neighbors and the effective values are governed by the details of the electronic structure. In order to examine which of the general conclusions from the modelling within the Heisenberg interactions can be reproduced from these effects of the electronic structure, we additionally perform weak-coupling pairing calculations.

\label{ap_sp_fl}
For this, we employ the well-established multiband spin-fluctuation pairing mechanism to our two-band electronic structure in absence of SLP to calculate the pairing interaction and then solve for the superconducting instability when allowing the eigenstates on the Fermi surface to pick up SLP. By this method, we exclude effects originating from changes in the Fermi surface geometry and topology that by themselves introduce transitions between superconducting instabilities \cite{Kreisel2026} and in general can lead to rich phase diagrams as discussed in view of unconventional superconductivity \cite{Romer2015, Scalapino2012}. In detail, we start from the normal state Hamiltonian, Eq.~(\ref{eq_normal_state}) together with the Hubbard interaction, i.e. the first term in Eq.~(~\ref{eq: Heisenberg interaction}) — and follow the multiband version of the spin-fluctuation pairing \cite{Graser2009,Scalapino2012,Kreisel2017} by calculating the non-interacting susceptibility tensor in the paramagnetic state,
\newcommand{\mbf}{\mathbf}
\renewcommand{\k}{{\mbf k}}
\newcommand{\q}{{\mbf q}}
 \begin{align}
 	\chi_{a b c d}^{(0)} (q) & = - \sum_{k,\alpha\beta} M_{abcd}^{\alpha\beta} (\k,\q)
 	 G^{\alpha}(k+q) G^{\beta} (k),   \label{eqn_supersuscept}
 \end{align}
where we have adopted the shorthand $k\equiv (\k,\omega_n)$, and  defined
\begin{equation}
	M_{a b c d}^{\alpha\beta} (\k,\q) = \unitary_{\beta d} (\k) \unitary_{ \beta b}^* (\k)
	\unitary_{\alpha a} (\k+\q) \unitary_{\alpha c}^* (\k+\q)
\end{equation}
as the products of $\unitary_{a \alpha}$ which is the matrix element of sublattice $a$ in the eigenstate with energy $\xi_\alpha(\k)$ of band $\alpha$.
In the band basis the  Green's function is given by
\begin{align}\label{eq_Gorb}
G^{\alpha} (\k,\omega_n)=  \frac{1}{\ii \omega_n -  \xi_\alpha(\k)}.
\end{align}
To include local interactions in a Hubbard model, i.e. the first term of Eq.~(\ref{eq: Heisenberg interaction})
\begin{align}
	H =  {U}\sum_{\textbf{R},a}
    n_{\uparrow a}(\textbf{R}) n_{\downarrow a}(\textbf{R})
	, \label{H}
\end{align}
we use the random phase approximation as follows.
First, the Matsubara sum in Eq.~(\ref{eqn_supersuscept}) is performed  analytically, and we then evaluate the relevant elements of $\chi_{a b c d}^0$ by integrating over the full Brillouin zone.
The spin- ($\chi_1^{\rm RPA}$) and charge-fluctuation ($\chi_0^{\rm RPA}$) parts of the RPA susceptibility for \mbox{$q=(\q,\omega_n=0)$} are now defined within the random phase approximation as
\begin{subequations}
\label{eqn:RPA}
\begin{align}
 \chi_{1\,abcd}^{\rm RPA} (\q) &= \left\{ \chi^{(0)} (q) \left[1 -\bar U^s \chi^{(0)} (q) \right]^{-1} \right\}_{abcd},\\
 \chi_{0\,abcd}^{\rm RPA} (\q) &= \left\{ \chi^{(0)} (q) \left[1 +\bar U^c \chi^{(0)} (q) \right]^{-1} \right\}_{abcd}.
\end{align}
\end{subequations}

The interaction matrices $\bar U^s$ and $\bar U^c$ in sublattice space contain the Hubbard parameter
$U$
and are just a simplification as already discussed in the multiorbital description \cite{Graser2009,Kreisel2017}.
The vertex for pair scattering between bands $\beta$ and $\alpha$ reads
\begin{equation} \label{eq:Gam_ij}
    \begin{aligned}
        	{\Gamma}_{\beta\alpha} (\k,\k') = & \mathrm{Re}
    \Big[\sum_{abcd} \unitary_{\beta a}^*(\k) \unitary_{\beta d}^*(-\k) \\
	 &\times {\Gamma}_{abcd} (\k,\k') \; \unitary_{\alpha b}(\k') \unitary_{\alpha c}(-\k') \Big]\,,
    \end{aligned}
\end{equation}
where $\k$ and $\k'$ are quasiparticle momenta restricted to the pockets $\k \in \rm{FS}^{(\beta)}$ and $\k' \in \rm{FS}^{(\alpha)}$. Here, the sublattice polarization enters our calculation via the modified components of the eigenvectors $U_{\alpha a}$. However, the vertex function in sublattice space is kept to the value without sublattice polarization given by
\begin{equation} \label{eq:fullGamma}
    \begin{aligned}
        	&{\Gamma}_{abcd} (\k,\k') = \left[\frac{3}{2} \bar U^s \chi_1^{\rm RPA} (\k-\k') \bar U^s \right.\,\,\\
	&\,\left. +  \frac{1}{2} \bar U^s - \frac{1}{2}\bar U^c \chi_0^{\rm RPA} (\k-\k') \bar U^c + \frac{1}{2} \bar U^c \right]_{abcd}.
    \end{aligned}
\end{equation}
To obtain the leading superconducting instability within this framework, we set up the linearized gap equation
 \begin{align}\label{eqn:gapeqn}
-  \sum_\alpha\frac{1}{|\Omega_\BZ|}
            \oint_{\rm{FS}^{(\alpha)}}
            \frac{\dd  \ell}{|\Vec{v}_{\rm{F}}^{(\alpha)}|} \Gamma_{\beta\alpha}^{\singtrip}(\k,\k')  g_i^\singtrip(\k')=\lambda_i^\singtrip g_{i}^\singtrip(\k)
 \end{align}
by discretizing the Fermi surface in a contour (in two dimensions)
and solving for the leading eigenvalue $\lambda$ and corresponding eigenfunction $g(\k)$.
Here $\Vec{v}_{\text{F}}^{(\alpha)}(\k')$ is the Fermi velocity of band $\alpha$ and we projected into the singlet (s) and triplet (t) channels by
$\Gamma_{\beta\alpha}^\singtrip(\k,\k')=\frac 12[\Gamma_{\beta\alpha}(\k,\k')\pm \Gamma_{\beta\alpha}(\k,-\k')]$.
The eigenfunction $g_i^\singtrip(\k)$ for the leading eigenvalue then determines the symmetry and structure of the leading pairing gap $\Delta^\singtrip(\k)\propto g_i^\singtrip(\k)$  close to $T_{\rm{c}}$ which is calculated from typically $\approx 200$ points on all Fermi sheets, i.e. by finding eigenvalues and eigenvectors of a $200\times 200$ matrix.
To get insights about the dominant pairing interactions, we perform a Fourier transform of the pairing vertex
\begin{align}
  \Gamma^{ab}_{\mathbf R} =\sum_{\mathbf k} {\Gamma}_{abba } (\k,-\k) {\rm{e}}^{-\ii \mathbf R\cdot\k},
\end{align}
which are presented in Figs.~\ref{fig: RPA case 1} and \ref{fig: RPA case 2}.

\bibliography{references}

\end{document}